\documentclass[%
 reprint,
preprintnumbers,
nofootinbib,
 amsmath,amssymb,
 prd,
]{revtex4-2}

\usepackage{graphicx}
\usepackage{dcolumn}
\usepackage{bm}
\usepackage{booktabs}
\usepackage{xcolor}
\usepackage{slashed}
\usepackage{amssymb}
\usepackage{lipsum}
\usepackage{multirow}

\begin{document}

\preprint{UCI-TR-2022-15}

\title{Entering the Era of Measuring Sub-Galactic Dark Matter Structure: \\
Accurate Transfer Functions for Axino, Gravitino \& Sterile Neutrino Thermal Warm  Dark Matter}

\author{Cannon M.\ Vogel}
\email{cvogel1@uci.edu}
\affiliation{Department of Physics and Astronomy,  University of California, Irvine, CA 92697-4575, USA}

\author{Kevork\ N.\ Abazajian}%
 \email{kevork@uci.edu}
\affiliation{Department of Physics and Astronomy,  University of California, Irvine, CA 92697-4575, USA}

\date{\today}

\begin{abstract} 
We examine thermal warm dark matter (WDM) models that are being probed by current constraints, and the relationship between the particle dark matter spin and commensurate thermal history. We find significant corrections to the linear matter power spectrum for given thermal WDM particle masses. Two primary classes are examined: spin-1/2 particles (e.g., thermalized sterile neutrinos, axinos) and thermal spin-3/2 particles (e.g., gravitinos or non-supersymmetric particles). We present new transfer function fits for thermal WDM candidates in particle mass regimes beyond the range of previous work, and at the scales of current and upcoming constraints. Importantly, we find that the standard, predominantly used, spin-1/2, thermal WDM particle produces a colder transfer function than that determined in previous work. We also analyze the entropy requirements for these WDM models to successfully produce observed dark matter densities. We explore the early Universe physics of gravitinos as either partially thermalized or fully thermalized species, which considerably changes the particle dark matter candidates' thermalization history and effects on structure formation. For the first time, we also calculate the transfer function for thermal spin-3/2 WDM.

\end{abstract}

\maketitle


\section{\label{sec:level1}Introduction}
Despite overwhelming evidence for its existence and detailed measurements of its abundance, the composition of dark matter (DM) is still unknown \cite{bertone2005particle, feng2010dark}. Cosmological and astrophysical methods of identification provide a testing ground for otherwise inaccessible theories; in particular, the velocity dispersion of dark matter is critical for determining structure formation and the resulting matter power spectrum which can be measured via clustering of galaxies and gas in the linear to mildly non-linear regime \cite{Colombi:1995ze, abazajian2006linear}. While the theory of a universe dominated by a cosmological constant and cold dark matter ($\Lambda$CDM) is highly successful---and hot dark matter is well ruled out---non-negligible velocity dispersions, corresponding to keV-scale particles, represent a viable region of parameter space that were originally motivated to solve challenges to the $\Lambda$CDM paradigm \cite{bode2001wdm,Bullock:2017xww}. Such warm dark matter (WDM) models reduce and eventually eliminate smaller mass halos in which galaxies can form, and also reduce the central densities in galaxy formation.

The decrease in power on small scales that results from a different free-streaming length in WDM occurs at a cutoff corresponding roughly to the scale of dwarf galaxies in the initial work on WDM \cite{bode2001wdm,lovell2014properties}. Above the free-streaming scale, WDM power matches CDM predictions, producing a targeted solution to a variety of problems arising in CDM models. One example is the interest in a thermal WDM particle mass scale of $\sim$2 keV as a solution to the central density (``Too Big to Fail") problem, as well as satellite counts, and distribution \cite{lovell2012haloes, anderhalden2013hints}. However, newer constraints from galaxy counts \cite{Polisensky:2010rw,Cherry:2017dwu,Nadler:2019zrb}, strong lensing \cite{Gilman:2019nap}, stellar streams \cite{Banik:2019smi},  and the Lyman-$\alpha$ forest (e.g., \cite{Irsic:2017ixq}), as well as combinations of these constraints place lower bounds on  the thermal WDM particle mass at scales much greater than 2 keV \cite{Banik:2019smi,Nadler:2021dft,Gilman:2021sdr,Zelko:2022tgf}. The latest combined constraints from strong lensing and galaxy counts place the lower limit on thermal WDM particle masses at approximately $m_\mathrm{th} > 9.8\,\mathrm{keV}$ (95\% CL) \cite{Zelko:2022tgf}. Limits from stellar streams, combined with galaxy counts, and lensing gives $m_\mathrm{th} > 11\,\mathrm{keV}$ \cite{Banik:2019smi}.  Such particles, with free-streaming scales corresponding to $\lesssim 3\times 10^5\, M_\odot$ are near or below the regime of interest for directly impacting galaxy formation, at even dwarf galaxy mass scales ($\sim\!10^7\,M_\odot$) \cite{Abazajian:2006yn}. Astrophysical probes are planned to continue the exploration of the free-streaming scale to ever-decreasing astrophysical mass scales, which correspond to increasing particle masses for thermal dark matter, ({\it viz.}, with JWST \cite{nierenberg2021definitive,Moustakas:2009na}). These constraints have largely considered only spin-1/2 thermal WDM.

Since current and future constraints are approaching increasingly small scales and higher precision and accuracy, a new examination of the connection between the linear structure formation in WDM models and their particle physics models is warranted, and the goal of this work. Specifically, what is often dubbed ``thermal WDM" is presented ambiguously in a particle-physics sense. The earliest work by \citet{bode2001wdm} and \citet{pagels1982grav} motivate and consider gravitino dark matter, a spin-3/2 particle, but specify their models to include a spin-1/2 particle. This has led to confusion in some parts of the literature, and will be explained in Section~\ref{sec:2}. While the massive gravitino is often best considered as having two spin degrees of freedom that are thermalized in the early Universe, other particle DM models allow for full thermalization. So, we also present the fully-thermalized spin-3/2 transfer functions.

Structure formation at small scales, $k\gtrsim 500\,h \,\mathrm{Mpc}^{-1}$, is also affected by new linear and nonlinear physics, at or near the smallest scale cutoffs we consider in this paper. These effects include the initial streaming velocity of baryons relative to the dark matter \cite{Tseliakhovich:2010bj}, and suppression of the baryon density fluctuations relative to dark matter even in the linear regime, due to baryonic pressure \cite{Naoz:2006tr}. Small-scale structure disruption is similarly important, with a range of halo masses between the filtering scale \cite{Naoz:2007fo} and Lyman-$\alpha$ cooling scale (virial temperature above $10^4 \, \mathrm{K}$) \cite{Oh:2002se}, corresponding to star-formation-suppressed halos with masses $0.2 \lesssim M_6 \lesssim 30$, where $M_6$ is halo mass in units of $10^6 \, M_\odot$. In this regime, higher halo densities relative to the intergalactic medium (IGM) can trap the ionization front, causing the gas component to explode in a sound crossing time, leading to nonlinear structure effects \cite{Hirata:2017ivs}. 
These effects would need to be included in analyses that rely on the smallest scale linear structure that we probe in this paper, but are beyond the scope of this work.

In the first part of this paper, we describe the particle physics in the early Universe of a variety of thermal WDM candidates, including gravitinos, axinos, and thermalized sterile neutrinos; and how that early period connects to linear growth of large to small-scale structure. We then describe how to implement these models' particle distributions in modern structure formation Boltzmann plus gravity solvers. We provide new results for the linear matter power at the $\lesssim$4\% level, and new fitting function forms for thermal WDM transfer functions, improving on previous fits. We also provide, for the first time, transfer functions for fully-thermalized spin-3/2 particles which diverge from partially thermalized gravitinos at the factor of $\sim$2 level.

\section{\label{sec:2}Thermal WDM Models}

\subsection{Spin-1/2}

Spin-1/2 thermal WDM is the most commonly considered WDM particle in observational studies. A variety of WDM particle models exist which remain viable as all or part of the dark matter. We consider here models for thermal WDM that constitute all of the dark matter. At a basic level, the number of spin degrees of freedom of the dark matter particle determines its abundance or density at its decoupling, while the change in entropy of the background plasma from decoupling until today affects the dark matter's dilution to its present abundance. Therefore, the spin state of the particle impacts the matter power spectrum which is eventually produced. One class of models consists of spin-1/2 particles, which includes both fully-thermalized sterile neutrinos and axinos.\footnote{Importantly, Dodelson-Widrow \cite{dodelson1994sterile} and Shi-Fuller \cite{shi1999new} sterile neutrino models are not in this class, as they never become fully thermalized in achieving the proper dark matter density.}

The axino ($\tilde{a}$) is the fermionic component of the axion supermultiplet in a supersymmetric (SUSY) model which incorporates an axion to solve the strong CP problem \cite{wilczek1991cosmological}. In the presence of supersymmetry breaking, the axino ceases to be degenerate and receives a mass contribution of order
\begin{equation}
    m_{\tilde{a}}\approx \mathcal{O}\left(\frac{m_{\mathrm{SUSY}}^2}{F_{\mathrm{PQ}}}\right)\, ,
\end{equation}
which was originally taken at the keV scale for the reasonable values, at the time, of $m_{\mathrm{SUSY}} \approx 10^3\, \mathrm{GeV}$ and $F_{\mathrm{PQ}} \approx 10^{12}\,\mathrm{GeV}$ \cite{wilczek1991cosmological}. More detailed modern analyses also find that a keV-scale axino is consistent with SUSY constraints under certain accidental symmetries that forbid leading order mass terms \cite{chun1992axino,kim2012mixing}. At this mass, the axino could be the lightest supersymmetric particle (LSP) which is stable on cosmological scales and decouples early after a period of full thermalization, thus being a viable WDM candidate. As such, the simple arguments from linear theory presented in Section~\ref{sec:3} produce an accurate picture of the resultant matter power spectrum. 

Sterile neutrinos may also fully thermalize through a new coupling, either related to the neutrino mass-generation mechanism or not. Such a coupling can determine sterile neutrinos' freeze-out at early times and provides a fully \textit{thermal} sterile neutrino dark matter candidate. One developed model that can achieve this is described by Jaramillo \cite{Jaramillo:2022mos}, which allows sterile neutrino thermalization from their Yukawa coupling prior to the electroweak transition, but suppresses the Yukawa after the transition so that the sterile neutrino is cosmologically stable. The nature of the coupling, freeze-out, and subsequent dilutions is not important in determining the structure formation effects of thermal sterile neutrino dark matter, as long as these quantities achieve the proper dark matter density, as described in \S\ref{sec:3}.

\subsection{Gravitinos and Other Spin-3/2 DM}

In addition to the axino, another potential LSP candidate for WDM is the gravitino. The gravitino ($\tilde{G}$) is the spin-3/2 SUSY partner of the graviton. In the symmetric supermultiplet, the gravitino is massless and thus can only access the $\pm$3/2 helicity states. As in the case of the axino, the gravitino can gain a mass via a super-Higgs process, ``eating" the goldstino, thus allowing the gravitino to access its $\pm$1/2 helicity states. If the theory is constructed in order to avoid introducing an excessive cosmological constant, the precise mass value becomes
\begin{equation}
    m_{\tilde{G}} = \frac{\kappa d}{\sqrt{6}} = \sqrt{\frac{4\pi}{3}}\frac{\Lambda_{SS}^2}{M_{Pl}} \, ,
\end{equation}
where $d$ is related to the SUSY-breaking scale $\Lambda_{SS} = \sqrt{d}$ and $\kappa/2$ is the gravitino coupling to the conserved vector-spinor current, which is given as $\kappa = \sqrt{8\pi}/M_{Pl}\approx 4\times10^{-19} \, \mathrm{GeV}^{-1}$ \cite{fayet1977mixing}. The gravitino mass is still effectively a free parameter, as $d$ is related to the SUSY-breaking scale; however, because $m_{\tilde{G}}$ is proportional to $\kappa$, and thus the inverse Planck mass, its presence in the polarization tensor allows for different interaction scales for different spin modes. While typical keV-mass scale gravitino interactions are weak to the point of cosmological insignificance \cite{choi1999cosmological}, the polarization tensor in this regime handily distinguishes the $\pm$1/2 helicity state components. Under the restriction of gravitino thermalization at energies significantly higher than its mass, the polarization tensor splits to become \cite{bolz2001thermal}
\begin{equation}
    \Pi_{\mu \nu} \approx -\slashed{P}g_{\mu \nu} +\frac{2}{3}\slashed{P}\frac{P_{\mu}P_{\nu}}{m_{\tilde{G}}^2}\, .
\end{equation}
This form comprises a tangential ($\pm$3/2) and longitudinal ($\pm$1/2) mode, respectively, where the longitudinal component is proportional to the inverse of the gravitino mass-squared, and thus proportional to the Planck mass-squared. As a result, the longitudinal mode's presence in a diagram provides a contribution of roughly unity, and thus they are comparatively easily thermalized in the early Universe \cite{fayet1977mixing}. This is the origin of the treatment of gravitino WDM as a goldstino; under the applicable assumptions for standard SUSY WDM, the two are effectively equivalent \cite{casalbuoni1988gravitino}. 

While these factors lead to the treatment of gravitinos as effectively spin-1/2 particles in most cases, a fully thermalized state of spin-3/2 dark matter is still of potential interest in model building which include mechanisms for thermalizing the transverse states in more general models, including those beyond supersymmetry \cite{Garcia:2020hyo,Ballesteros:2020adh}. Additionally, accurate transfer functions for fully-thermalized WDM at spin-3/2, which behave more closely to CDM and are thus more difficult to constrain, provide a tool by which future precision cosmology searches could distinguish between models. Quantitatively, spin-3/2 states have colder transfer functions. The half-modes of the spin-3/2 transfer functions, where $T_X(k)=1/2$, are at $k$-scales 16\% to 20\% larger than the spin-1/2 case (i.e., colder).

\begin{figure*}[ht!]
\centering
\includegraphics[width=0.48\textwidth]{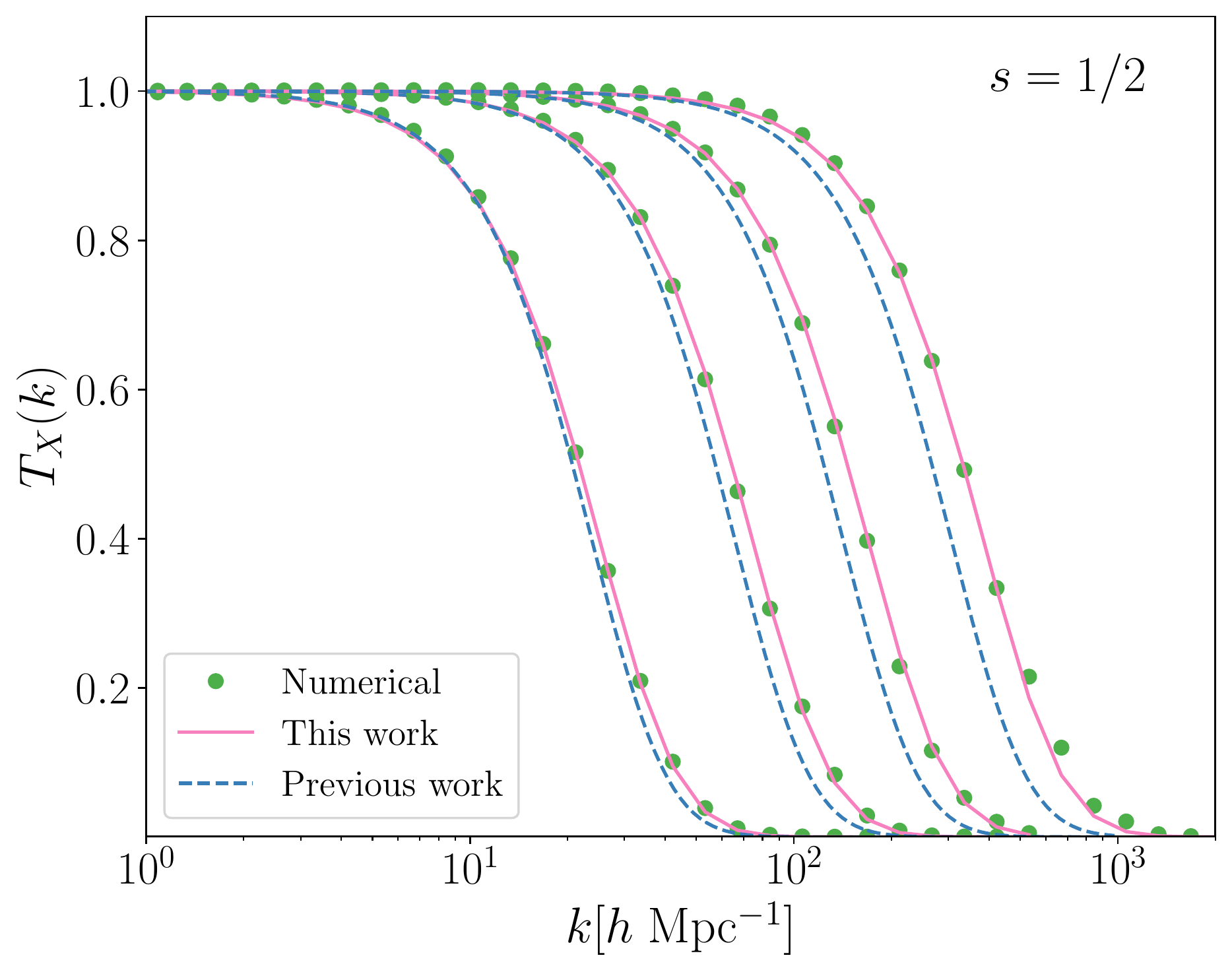}
\includegraphics[width=0.48\textwidth]{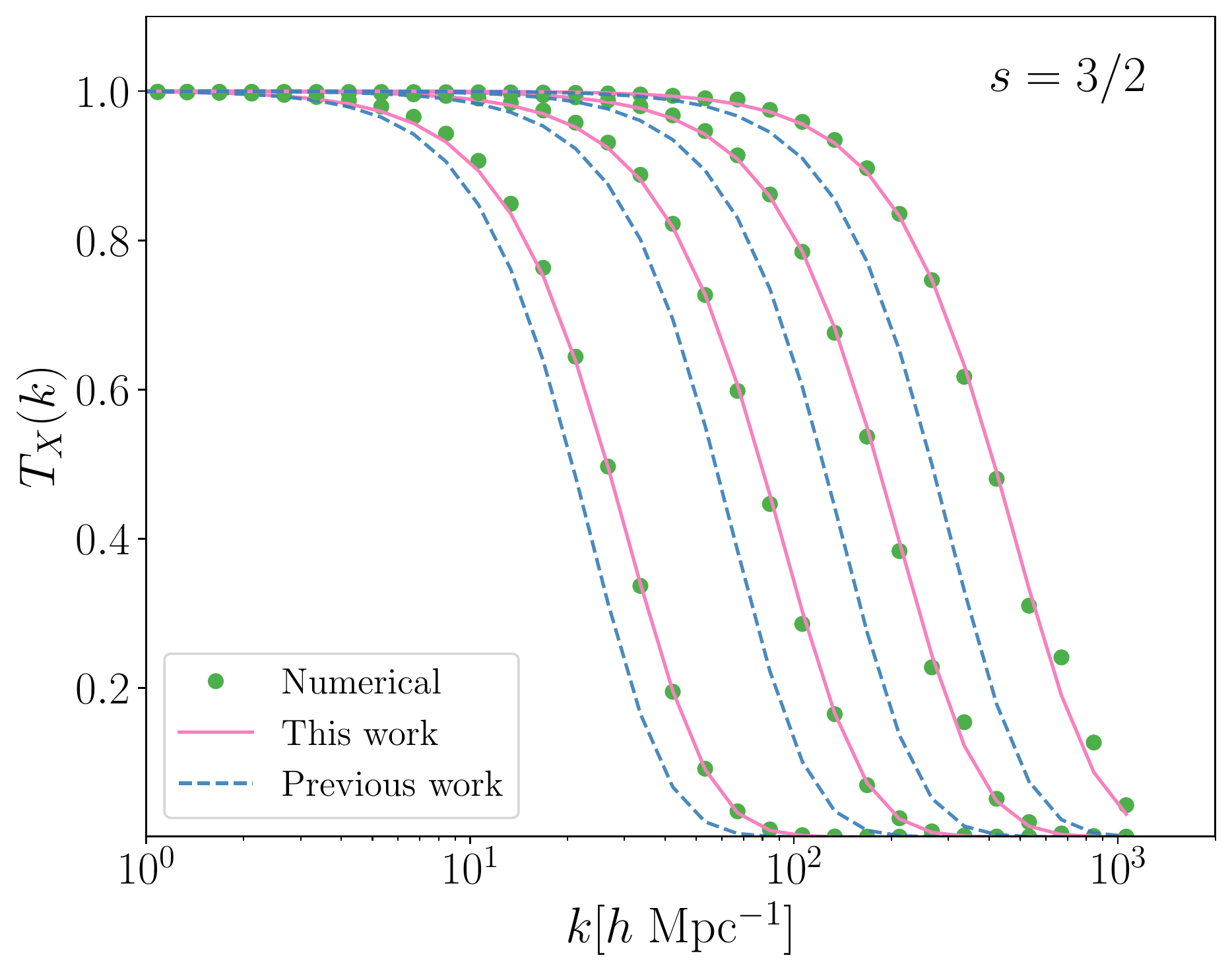} 
\caption{The transfer functions for fully thermalized spin-1/2 (left) and spin-3/2 (right) WDM. From left to right are masses 2, 5, 10, and 20 keV. The green data points represent the results of CLASS numerical calculation; the solid pink lines are our best fits; and the dashed blue lines are the fits from Viel \emph{et al.} \cite{viel2005}, which are appropriate to use only for the spin-1/2 case, but we show them here for comparison.\label{fig:transfer}
}
\end{figure*}

\begin{figure}[ht!]
\centering

\includegraphics[width=0.48\textwidth]{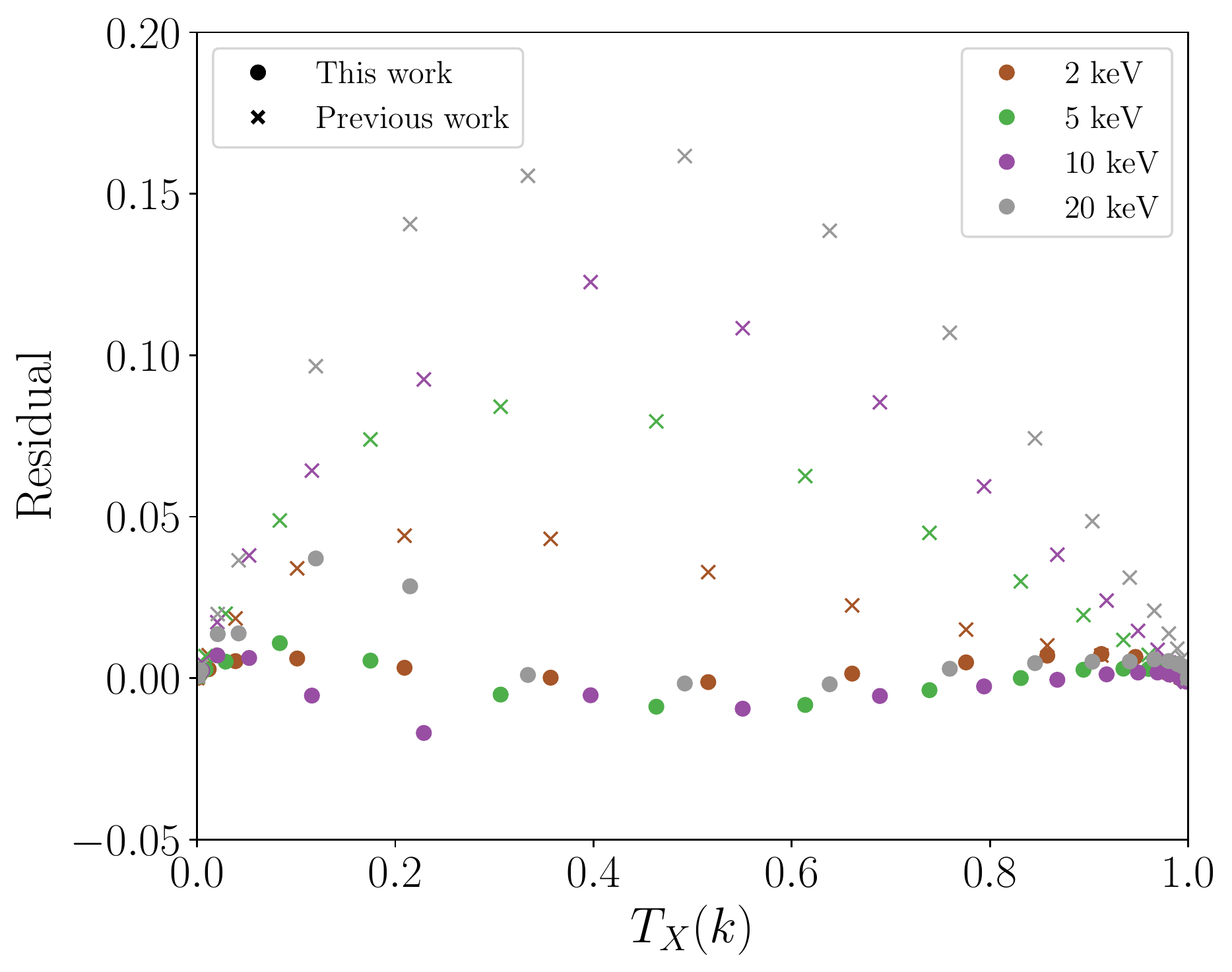}
\caption{The transfer function residuals for fully thermalized spin-1/2 WDM. Each color represents a distinct mass value, as shown in the legend. The circles ($\bullet$) are our fits' residual values, which are within $4\%$ of the numerical calculations, and crosses ($\times$) are Viel \emph{et al.}~\cite{viel2005} residual values, which are up to $16\%$ of the numerical calculations for the 20 keV particle mass scale. How this maps onto an inferred particle-mass is nonlinear, and provided in Fig.~\ref{fig:mvsm}. \label{fig:1/2-res}
}
\vskip -0.5 cm
\end{figure}

\section{\label{sec:3}The Matter Power Spectrum and WDM}

In order to analyze the power spectra of WDM candidates, it is important to accurately account for the results of linear theory with consideration of the background entropy. For the standard neutrino, equating the scaled total entropy density before and after electron-positron annihilation gives the canonical relative temperature of $T_\nu/T_\gamma = (4/11)^{1/3}$. For the case of a WDM particle, $X$, which was once in thermal equilibrium, the relative temperature is found in a similar way to the case of neutrinos, with an extra factor from the dilution due background particle annihilation into photons, electrons, positrons, and neutrinos ($g_\ast=43/4$), giving a ratio 
\begin{equation}
    \frac{T_{X}}{T_\gamma} = \left(\frac{4}{11}\right)^{1/3}\left(\frac{43/4}{g_{*}(T_D)}\right)^{1/3}.
\end{equation}

In order to find the spin degrees of freedom in the plasma at decoupling, $g_*(T_D)$, for a given model, one must find a relative abundance in terms of the degrees of freedom (DoF), $g_X$, which can then be matched to measurements of dark matter density. The WDM-photon ratio is a simple product of the dilution of WDM particles relative to photons when the original DoF convert into photons, electrons, positrons, and neutrinos at temperature $T_D$; the electron-positron annihilation dilution; and the relative DoF to photons, respectively \cite{bode2001wdm},  
\begin{equation}
    \frac{n_{X}}{n_\gamma} = \left(\frac{43/4}{g_*(T_D)}\right)\left(\frac{4}{11}\right)\left(\frac{g_{X}}{2}\right)\, .
\end{equation}
Once the relative abundance is found, the relation $\rho = m_{X}n_{X}$ immediately gives the density as

\begin{align}
    \omega_X &\equiv \Omega_X h^2  = \frac{43}{11}\frac{\zeta_3T^3h^2}{\pi^2\rho_\mathrm{cr}}\frac{g_X}{g_*(T_D)}m_\mathrm{th}\nonumber\\
    &\approx \frac{115}{g_*(T_D)}\frac{g_X}{1.5}\frac{m_\mathrm{th}}{\mathrm{keV}}    \, ,
    \label{eq:omega}
\end{align} 
where $\rho_\mathrm{cr}$ is the critical density, and $\zeta_3$ is the Riemann zeta function evaluated at 3.

With this, one can simply supply a WDM density and find both the DoF at decoupling and the relative temperature of the WDM. Table~\ref{tab:grav_params}  contains example values for fully-thermalized WDM models under standard cosmological parameters, where the WDM particle is the entire DM. Importantly, a very large early entropy is required for gravitino WDM to comprise all of the dark matter. Spin-1/2 dark matter particles have similar dilution requirements. While these values are quite large relative to the standard model predictions at early times, various entropy producing mechanisms may alleviate such concerns, through many degrees of freedom in the particle content of the model, or via intervening phase transitions \cite{baltz2003gravitino}. 

\begin{table}[t]
  \begin{center}
    \begin{tabular}{c|c|c|c|c}
      \toprule\midrule[0.3pt]
      \ & \multicolumn{2}{c|}{Spin-1/2} &\multicolumn{2}{c}{Spin-3/2}  \\
      
      $m$ [keV] & $g_*(T_D)$ & $T_X/T_\gamma$&$g_*(T_D)$ & $T_X/T_\gamma$\\
      \midrule 
      2 & 1917 & 0.1268 & 3833 & 0.1007  \\
      5 & 4792 & 0.09344 & 9583 & 0.07416 \\
      10 & 9583 & 0.07416 & 19170 & 0.05886\\
      20 &  19170 & 0.05886 & 38330 & 0.04672 \\
      \midrule[0.3pt]\bottomrule 
    \end{tabular}
  \end{center}
    \caption{Example background plasma entropy and resulting temperatures for fully-thermalized spin-1/2 and spin-3/2 WDM. We note the clear mass-spin space relation as characterized by Eq.~\eqref{eq:omega}.    \label{tab:grav_params}}
\end{table}

Once the required relative temperature of the WDM to the photons is known, it can be used along with an Einstein-Boltzmann solver to find the matter power spectrum. This power spectrum takes a form which mimics that for CDM, only diverging at small scales with a suppression of power \cite{viel2012non}. In order to conveniently encode the relative spectral form, the WDM transfer function is defined as
\begin{equation}
    T_X(k) \equiv \sqrt{\frac{P(k)_{\Lambda\mathrm{WDM}}}{P(k)_{\Lambda\mathrm{CDM}}}}\, ,
\end{equation}
for the same cosmology. The function is normalized to unity at large scales and has been historically parameterized with the function
\begin{equation}
    T_X(k;\alpha,\nu) = [1+(\alpha k)^{2\nu}]^{-5/\nu}\, ,
    \label{eq:transfit}
\end{equation}
where $\alpha$ and $\nu$ are model and cosmology dependent parameters derived from numerical calculation \cite{bode2001wdm, viel2005,hansen2002constraining,abazajian2006production}. Once a model and cosmology are specified, the above method provides a path to produce a relative measure of the matter power spectrum applicable to future studies and simulations.

\begin{figure*}[t]
\centering
\includegraphics[width=0.48\textwidth]{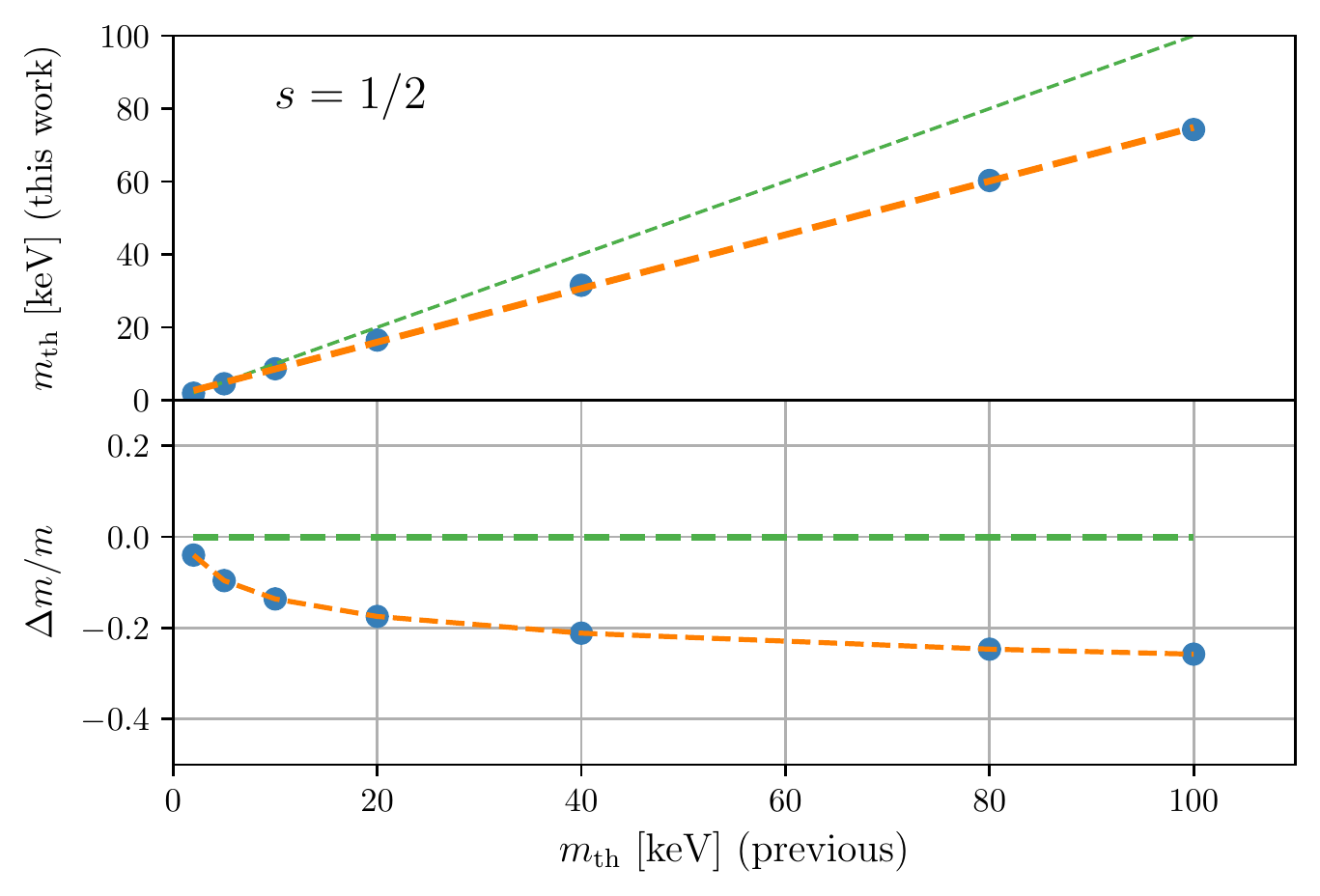}\includegraphics[width=0.48\textwidth]{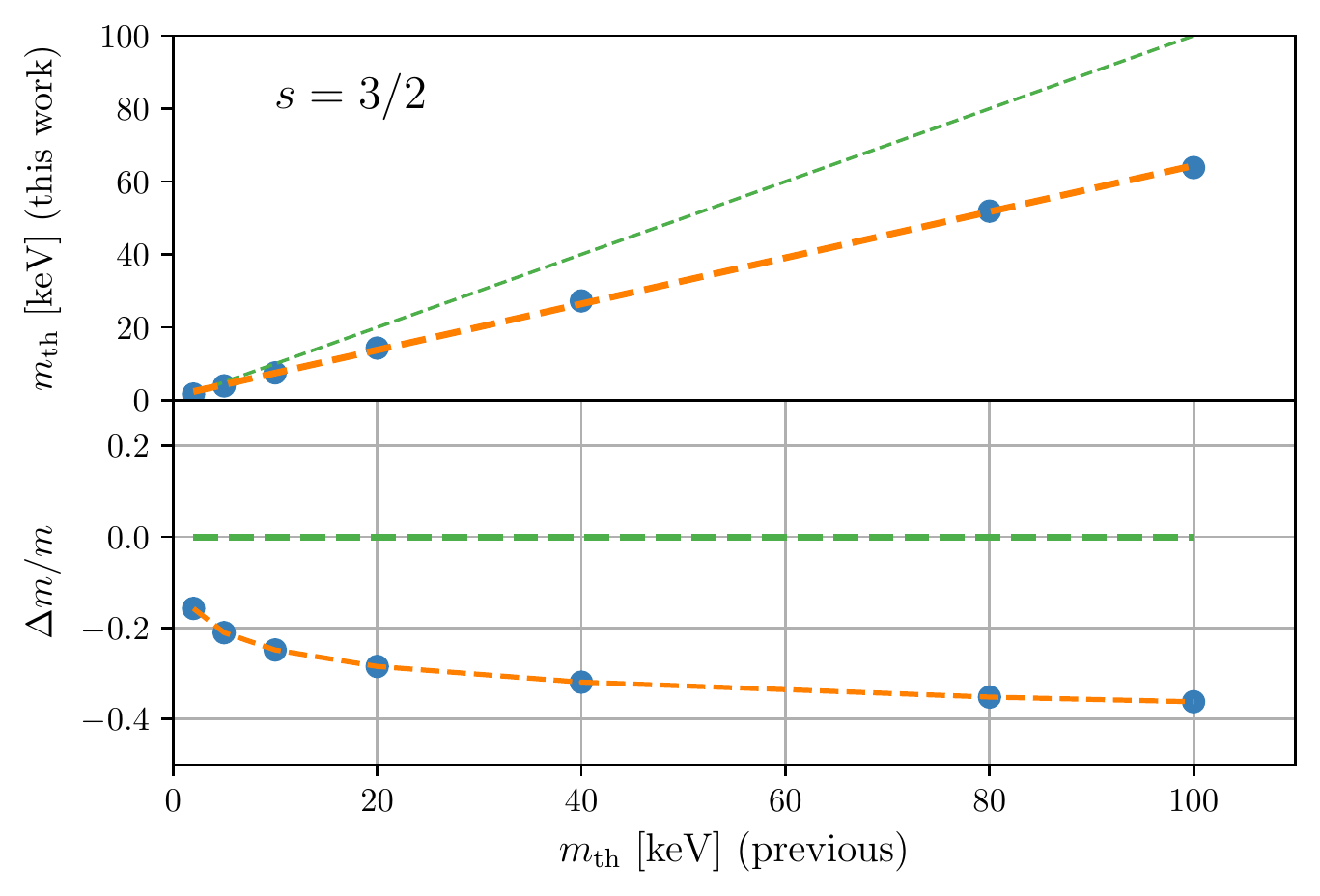}

\caption{ The thermal particle mass from our fits in relation to the previous fits for spin-1/2 (left) and spin-3/2 (right) thermal WDM. The lower panels show the fractional change, demonstrating that the new fits lead to significantly improved mass matching as we move to smaller scales and higher particle masses. The green line in all panels represents no change. 
\label{fig:mvsm}}
\end{figure*}

\section{\label{sec:4}Methods \& Results}

In order to calculate the linear density fluctuations which give rise to the matter power spectrum, we use the CLASS Boltzmann code \cite{blas2011cosmic}. The benchmark parameters for the $\Lambda$CDM model are taken from Planck 2018 \cite{aghanim2020planck}. For accuracy, we include the Planck 2018 minimal case of one massive neutrino species with $m_\nu = 0.06\,\mathrm{eV}$,  modeled as a non-CDM (NCDM) dark matter component in CLASS. Specifically, we adopt the cosmological parameters that are the best estimate parameters from the TT,TE,EE+lowE+lensing column in Table 2 of Ref.~\cite{aghanim2020planck}, and we provide a fitting formula below for cosmological parameters within Planck 2018 stated $2\sigma$ uncertainties, generated via multi-dimensional fitting to a large set of CLASS runs with varying cosmologies. For the pure WDM models we consider here, CDM density is replaced by WDM, which in CLASS is handled as NCDM. To fully specify the WDM model, we then only need to provide a particle mass and the relative temperature, as described in Section~\ref{sec:3} and given in Table~\ref{tab:grav_params}. Multiple runs were performed for our highest mass case, 100 keV, which extends to the highest wavenumber, or smallest scales, where numerical effects are most prevalent. We tested different NCDM tolerance values ranging from $10^{-2}$ to $10^{-7}$ (specified by \texttt{tol\_ncdm\_bg} and \texttt{tol\_ncdm\_gauge}), which minimally affected our fit results. We adopted a value of $10^{-3}$ to optimize computational efficiency and accuracy. Gauge choice of the NCDM made no significant differences.  

For this analysis, we analyzed fully thermalized, spin-1/2 and spin-3/2 particles at masses of 2, 5, 10, 20, and 100 keV. The resulting matter power spectra were then fit to the transfer function using the functional form in Eq.~\eqref{eq:transfit}, under a specific form of $\alpha$,
\begin{equation}
    \alpha = a\, m_\mathrm{th}^{b} \left(\frac{\omega_X}{0.12}\right)^{\eta}\left(\frac{h}{0.6736}\right)^{\theta} h^{-1}\, \mathrm{Mpc} \, .
\end{equation}
We use a non-linear, least-squares method in mass-wavenumber space to fit the three free WDM-dependent parameters ($a$, $b$, and $\nu$). We also included the case where the powers in Eq.~\eqref{eq:transfit} are independent, i.e., adding a new parameter, but the improvement of the fit is at the $10^{-4}$ level, much smaller than our remaining residual uncertainties. The best-fit transfer-function model was then used to find residuals with respect to CLASS as well as previous WDM transfer-function fits. 

To parameterize the dependence of our results on key cosmological parameters, $\omega_X$ and $h$ were separately varied for the case of the 2 keV, spin-1/2 particle, and the results were linearly fit to provide the cosmology dependence in $\alpha$. The cosmological dependence for spin-3/2 particle models was similar. In order to properly treat the reliance on cosmology, the CDM and WDM models used in the transfer function were both simultaneously varied, which, in the WDM case, requires the adjustment of the $g_*(T_D)$ and thus relative temperature calculations as described in Section~\ref{sec:3}. The results were included in the functional form of $\alpha$ via the parameters $\eta$ and $\theta$.


The results of the transfer function fits for each tested particle mass are shown for spin-3/2 particles and for spin-1/2 particles in Figure~\ref{fig:transfer}. The dashed comparison curves are taken from \citet{viel2005}, which analyzed spin-1/2 particles between $0.092$ keV and $1.441 \, \mathrm{keV}$, resulting in a poorer fit compared to our results in this higher particle mass regime. The residuals are plotted separately in Figure~\ref{fig:1/2-res} for the spin-1/2 model. The spin-3/2 model fit had similar residuals with respect to the numerical calculation. We provide the best fit values for all parameters in Table~\ref{tab:fit_params}. 

It can be seen in Figs.~\ref{fig:transfer} and \ref{fig:1/2-res} that even in the case of spin-1/2 thermal WDM, our transfer function fits differ from the previous work at all scales of wavenumber, $k$, where the transfer function deviates from unity. Therefore, all constraints that are claimed at the $\gtrsim 1\,\mathrm{keV}$ scale that rely on the linear matter transfer function to determine their constraints would need to be corrected in order to map onto the proper dark matter particle mass. Though different observables may rely on different portions of the transfer function, we match the previous fit with our newly derived forms at $T_X(k)=1/2$, a benchmark scale for the effect of WDM often used in previous work (cf. Ref.~\cite{Zelko:2022tgf}). 

In Fig.~\ref{fig:mvsm}, we show the level of the correction on the inferred particle mass when matching the previous fitting form (Viel et al.\cite{viel2005}) and our fit at $T_X(k)=1/2$, for both spin-1/2 and spin-3/2. Current constraints on thermal WDM focus on spin-1/2 particles. For spin-1/2 thermal WDM, the particle-mass correction goes from $14\%$ at $10\,\mathrm{keV}$, at the range of current constraints, to  $26\%$ at $100\,\mathrm{keV}$. The exact level of the correction on the observable's constraint on particle mass would depend on which portion of the transfer function that the observable is most sensitive. We encourage employment of our more accurate functions, since they differ from the previous work at all scales of wavenumber, $k$, where the transfer function deviates from unity, with even a $> 5\%$ correction for spin-1/2 10 keV transfer functions at 20\% suppressions (i.e., $T_X(k)=0.8$, see Fig.~\ref{fig:1/2-res}).

An important observation from this analysis is that the particle mass value observed in the fully-thermalized spin-3/2 model is linearly related to an effective spin-1/2 particle rest mass, going beyond the method of \citet{bode2001wdm} which limits the functional form of $\alpha$ to be a power law in terms of $g_x$. While this linear parameterization of $m$ in terms of $g_x$ may be useful, the limited applicability of higher spin models led us to not further study the functional form with CLASS runs for higher spin than 3/2. However, this implies that any future constraints based on WDM transfer functions may treat mass bounds as effectively mass-spin bounds, as the two produce the same shifting effect in the transfer function. 

\begin{table}[t]
  \begin{center}
    
    \begin{tabular}{c|c|c|c|c|c
    }
      \toprule \midrule[0.3pt]
      model & $a$ & $b$ & $\nu$ & $\theta$ & $\eta$ \\
      \midrule 
      spin-1/2 &0.0437 &-1.188  &1.049 & 2.012&0.2463 \\
      spin-3/2 &0.0345 &-1.195  &1.025 & -- & --\\
      \midrule[0.3pt]\bottomrule 
    \end{tabular}
  \end{center}
\caption{Parameter fit values for spin-1/2 and spin-3/2 WDM transfer functions, as described in the text.\label{tab:fit_params}}

\end{table}

\section{\label{sec:6}Conclusions}  
In this paper, we explored the early Universe physics of spin-1/2 and spin-3/2 thermal WDM particle models, which differ substantially. We calculate, for the first time, the transfer functions for thermal spin-3/2 WDM. We further calculated cosmological linear structure formation given by the matter power spectrum up to particle masses of $100\,\mathrm{keV}$, a range relevant for current and forthcoming structure formation observational sensitivities. Importantly, our updated standard, spin-1/2, thermal WDM transfer function is colder than that from previous work, which is used predominantly in the literature. This deviation is present at all wavenumbers $k$ for which the transfer function deviates from unity. It deviates the most near $T_X(k)\approx 1/2$, but not exactly there for all particle masses (cf. Fig.~\ref{fig:1/2-res}). To give a sense of scale to the corrections, we match previous forms of thermal WDM transfer functions at a suppression scale of $T_X(k)=1/2$, as used in previous work. Employing this matching scale, our fitting forms for the canonical spin-1/2 thermal WDM particle correct the inferred particle mass by 14\% at 10 keV and 26\% at 100 keV. We also show that the corrections on the transfer function are significant at small suppressions (at smaller $k$). Therefore, we strongly encourage the use of our more accurate thermal WDM transfer functions for constraints that use any part of the thermal WDM transfer function. This would be particularly important if a detection of a thermal WDM cutoff was found, leading to an inferred specific dark matter particle mass that could be searched for via other methods. 

Our more accurate forms are largely due to calibration at higher thermal WDM particle masses. Previous work fit to thermal WDM particle masses less than 1.5 keV \cite{viel2005}, which were the scales being probed at that time. Our transfer-function fits match CLASS-based numerical calculations to the 4\% level, providing more confidence in future analyses' investigations of WDM signatures and their mapping to early Universe particle physics. 

In our investigation, we calculate the entropy requirements for these WDM models, finding that high $g_\ast$ values in the early Universe are required, up to $g_\ast \sim 20,\!000$, for WDM to comprise all of the dark matter. This $g_\ast$ is certainly large, but it may be provided by the particle content of the model, or via intervening phase transitions. We specifically explore the difference between spin-3/2 and the canonical spin-1/2 thermal WDM thermal histories and effects on structure formation. As discussed earlier, spin-3/2 particle transfer functions are equivalent to a shifted spin-1/2 particle transfer functions, via a linear relation. Therefore, future constraints on WDM structure could be viewed as particle mass bounds in mass-spin space. Other nonlinear and linear effects at the smallest scales, {\it viz.} $k>500 h^{-1}\rm\, Mpc$, also modify the evolution of perturbations and halo growth, as described above, and should be considered when structure formation scenarios require detailed modeling of those scales.

Our work is relevant for determinations of the lower bounds on WDM particle masses which are approaching or at the $\sim 10\,\mathrm{keV}$ scale, including from galaxy counts \cite{Polisensky:2010rw,Cherry:2017dwu,Nadler:2019zrb}, the Lyman-$\alpha$ forest \cite{Irsic:2017ixq}, stellar streams \cite{Banik:2019smi}, and strong lensing \cite{Gilman:2019nap}, as well as combined constraints \cite{Zelko:2022tgf}. In particular, if a WDM free-streaming scale is inferred from observations, e.g. as weakly favored in recent Lyman-$\alpha$ forest analyses \cite{Villasenor:2022aiy}, our work will more accurately specify the particle dark matter properties involved as well as the history of the early Universe.

\acknowledgements 

CMV and KNA acknowledge useful conversations with Jo Bovy, Jonathan Feng, Alex Kusenko, Ethan Nadler, Anna Nierenberg, Devon Powell, Arvind Rajaraman, and Ioanna Zelko. We would also like to thank the organizers of the UCLA Dark Matter 2023 Conference, where many of these discussions occurred. KNA is partially supported by U.S. National Science Foundation (NSF) Theoretical Physics Program, Grants PHY-1915005 and PHY-2210283.

\bibliography{thermWDM}

\providecommand{\noopsort}[1]{}\providecommand{\singleletter}[1]{#1}%
\begin{thebibliography}{47}%
\makeatletter
\providecommand \@ifxundefined [1]{%
 \@ifx{#1\undefined}
}%
\providecommand \@ifnum [1]{%
 \ifnum #1\expandafter \@firstoftwo
 \else \expandafter \@secondoftwo
 \fi
}%
\providecommand \@ifx [1]{%
 \ifx #1\expandafter \@firstoftwo
 \else \expandafter \@secondoftwo
 \fi
}%
\providecommand \natexlab [1]{#1}%
\providecommand \enquote  [1]{``#1''}%
\providecommand \bibnamefont  [1]{#1}%
\providecommand \bibfnamefont [1]{#1}%
\providecommand \citenamefont [1]{#1}%
\providecommand \href@noop [0]{\@secondoftwo}%
\providecommand \href [0]{\begingroup \@sanitize@url \@href}%
\providecommand \@href[1]{\@@startlink{#1}\@@href}%
\providecommand \@@href[1]{\endgroup#1\@@endlink}%
\providecommand \@sanitize@url [0]{\catcode `\\12\catcode `\$12\catcode
  `\&12\catcode `\#12\catcode `\^12\catcode `\_12\catcode `\%12\relax}%
\providecommand \@@startlink[1]{}%
\providecommand \@@endlink[0]{}%
\providecommand \url  [0]{\begingroup\@sanitize@url \@url }%
\providecommand \@url [1]{\endgroup\@href {#1}{\urlprefix }}%
\providecommand \urlprefix  [0]{URL }%
\providecommand \Eprint [0]{\href }%
\providecommand \doibase [0]{http://dx.doi.org/}%
\providecommand \selectlanguage [0]{\@gobble}%
\providecommand \bibinfo  [0]{\@secondoftwo}%
\providecommand \bibfield  [0]{\@secondoftwo}%
\providecommand \translation [1]{[#1]}%
\providecommand \BibitemOpen [0]{}%
\providecommand \bibitemStop [0]{}%
\providecommand \bibitemNoStop [0]{.\EOS\space}%
\providecommand \EOS [0]{\spacefactor3000\relax}%
\providecommand \BibitemShut  [1]{\csname bibitem#1\endcsname}%
\let\auto@bib@innerbib\@empty
\bibitem [{\citenamefont {Bertone}\ \emph {et~al.}(2005)\citenamefont
  {Bertone}, \citenamefont {Hooper},\ and\ \citenamefont
  {Silk}}]{bertone2005particle}%
  \BibitemOpen
  \bibfield  {author} {\bibinfo {author} {\bibfnamefont {G.}~\bibnamefont
  {Bertone}}, \bibinfo {author} {\bibfnamefont {D.}~\bibnamefont {Hooper}}, \
  and\ \bibinfo {author} {\bibfnamefont {J.}~\bibnamefont {Silk}},\ }\href@noop
  {} {\bibfield  {journal} {\bibinfo  {journal} {Physics reports}\ }\textbf
  {\bibinfo {volume} {405}},\ \bibinfo {pages} {279} (\bibinfo {year}
  {2005})}\BibitemShut {NoStop}%
\bibitem [{\citenamefont {Feng}(2010)}]{feng2010dark}%
  \BibitemOpen
  \bibfield  {author} {\bibinfo {author} {\bibfnamefont {J.~L.}\ \bibnamefont
  {Feng}},\ }\href@noop {} {\bibfield  {journal} {\bibinfo  {journal} {Annual
  Review of Astronomy and Astrophysics}\ }\textbf {\bibinfo {volume} {48}},\
  \bibinfo {pages} {495} (\bibinfo {year} {2010})}\BibitemShut {NoStop}%
\bibitem [{\citenamefont {Colombi}\ \emph {et~al.}(1996)\citenamefont
  {Colombi}, \citenamefont {Dodelson},\ and\ \citenamefont
  {Widrow}}]{Colombi:1995ze}%
  \BibitemOpen
  \bibfield  {author} {\bibinfo {author} {\bibfnamefont {S.}~\bibnamefont
  {Colombi}}, \bibinfo {author} {\bibfnamefont {S.}~\bibnamefont {Dodelson}}, \
  and\ \bibinfo {author} {\bibfnamefont {L.~M.}\ \bibnamefont {Widrow}},\
  }\href {\doibase 10.1086/176788} {\bibfield  {journal} {\bibinfo  {journal}
  {Astrophys. J.}\ }\textbf {\bibinfo {volume} {458}},\ \bibinfo {pages} {1}
  (\bibinfo {year} {1996})},\ \Eprint {http://arxiv.org/abs/astro-ph/9505029}
  {arXiv:astro-ph/9505029} \BibitemShut {NoStop}%
\bibitem [{\citenamefont
  {Abazajian}(2006{\natexlab{a}})}]{abazajian2006linear}%
  \BibitemOpen
  \bibfield  {author} {\bibinfo {author} {\bibfnamefont {K.}~\bibnamefont
  {Abazajian}},\ }\href@noop {} {\bibfield  {journal} {\bibinfo  {journal}
  {Physical Review D}\ }\textbf {\bibinfo {volume} {73}},\ \bibinfo {pages}
  {063513} (\bibinfo {year} {2006}{\natexlab{a}})}\BibitemShut {NoStop}%
\bibitem [{\citenamefont {Bode}\ \emph {et~al.}(2001)\citenamefont {Bode},
  \citenamefont {Ostriker},\ and\ \citenamefont {Turok}}]{bode2001wdm}%
  \BibitemOpen
  \bibfield  {author} {\bibinfo {author} {\bibfnamefont {P.}~\bibnamefont
  {Bode}}, \bibinfo {author} {\bibfnamefont {J.~P.}\ \bibnamefont {Ostriker}},
  \ and\ \bibinfo {author} {\bibfnamefont {N.}~\bibnamefont {Turok}},\
  }\href@noop {} {\bibfield  {journal} {\bibinfo  {journal} {The Astrophysical
  Journal}\ }\textbf {\bibinfo {volume} {556}},\ \bibinfo {pages} {93}
  (\bibinfo {year} {2001})}\BibitemShut {NoStop}%
\bibitem [{\citenamefont {Bullock}\ and\ \citenamefont
  {Boylan-Kolchin}(2017)}]{Bullock:2017xww}%
  \BibitemOpen
  \bibfield  {author} {\bibinfo {author} {\bibfnamefont {J.~S.}\ \bibnamefont
  {Bullock}}\ and\ \bibinfo {author} {\bibfnamefont {M.}~\bibnamefont
  {Boylan-Kolchin}},\ }\href {\doibase 10.1146/annurev-astro-091916-055313}
  {\bibfield  {journal} {\bibinfo  {journal} {Ann. Rev. Astron. Astrophys.}\
  }\textbf {\bibinfo {volume} {55}},\ \bibinfo {pages} {343} (\bibinfo {year}
  {2017})},\ \Eprint {http://arxiv.org/abs/1707.04256} {arXiv:1707.04256
  [astro-ph.CO]} \BibitemShut {NoStop}%
\bibitem [{\citenamefont {Lovell}\ \emph {et~al.}(2014)\citenamefont {Lovell},
  \citenamefont {Frenk}, \citenamefont {Eke}, \citenamefont {Jenkins},
  \citenamefont {Gao},\ and\ \citenamefont {Theuns}}]{lovell2014properties}%
  \BibitemOpen
  \bibfield  {author} {\bibinfo {author} {\bibfnamefont {M.~R.}\ \bibnamefont
  {Lovell}}, \bibinfo {author} {\bibfnamefont {C.~S.}\ \bibnamefont {Frenk}},
  \bibinfo {author} {\bibfnamefont {V.~R.}\ \bibnamefont {Eke}}, \bibinfo
  {author} {\bibfnamefont {A.}~\bibnamefont {Jenkins}}, \bibinfo {author}
  {\bibfnamefont {L.}~\bibnamefont {Gao}}, \ and\ \bibinfo {author}
  {\bibfnamefont {T.}~\bibnamefont {Theuns}},\ }\href@noop {} {\bibfield
  {journal} {\bibinfo  {journal} {Monthly Notices of the Royal Astronomical
  Society}\ }\textbf {\bibinfo {volume} {439}},\ \bibinfo {pages} {300}
  (\bibinfo {year} {2014})}\BibitemShut {NoStop}%
\bibitem [{\citenamefont {Lovell}\ \emph {et~al.}(2012)\citenamefont {Lovell},
  \citenamefont {Eke}, \citenamefont {Frenk}, \citenamefont {Gao},
  \citenamefont {Jenkins}, \citenamefont {Theuns}, \citenamefont {Wang},
  \citenamefont {White}, \citenamefont {Boyarsky},\ and\ \citenamefont
  {Ruchayskiy}}]{lovell2012haloes}%
  \BibitemOpen
  \bibfield  {author} {\bibinfo {author} {\bibfnamefont {M.~R.}\ \bibnamefont
  {Lovell}}, \bibinfo {author} {\bibfnamefont {V.}~\bibnamefont {Eke}},
  \bibinfo {author} {\bibfnamefont {C.~S.}\ \bibnamefont {Frenk}}, \bibinfo
  {author} {\bibfnamefont {L.}~\bibnamefont {Gao}}, \bibinfo {author}
  {\bibfnamefont {A.}~\bibnamefont {Jenkins}}, \bibinfo {author} {\bibfnamefont
  {T.}~\bibnamefont {Theuns}}, \bibinfo {author} {\bibfnamefont
  {J.}~\bibnamefont {Wang}}, \bibinfo {author} {\bibfnamefont {S.~D.}\
  \bibnamefont {White}}, \bibinfo {author} {\bibfnamefont {A.}~\bibnamefont
  {Boyarsky}}, \ and\ \bibinfo {author} {\bibfnamefont {O.}~\bibnamefont
  {Ruchayskiy}},\ }\href@noop {} {\bibfield  {journal} {\bibinfo  {journal}
  {Monthly Notices of the Royal Astronomical Society}\ }\textbf {\bibinfo
  {volume} {420}},\ \bibinfo {pages} {2318} (\bibinfo {year}
  {2012})}\BibitemShut {NoStop}%
\bibitem [{\citenamefont {Anderhalden}\ \emph {et~al.}(2013)\citenamefont
  {Anderhalden}, \citenamefont {Schneider}, \citenamefont {Maccio},
  \citenamefont {Diemand},\ and\ \citenamefont
  {Bertone}}]{anderhalden2013hints}%
  \BibitemOpen
  \bibfield  {author} {\bibinfo {author} {\bibfnamefont {D.}~\bibnamefont
  {Anderhalden}}, \bibinfo {author} {\bibfnamefont {A.}~\bibnamefont
  {Schneider}}, \bibinfo {author} {\bibfnamefont {A.~V.}\ \bibnamefont
  {Maccio}}, \bibinfo {author} {\bibfnamefont {J.}~\bibnamefont {Diemand}}, \
  and\ \bibinfo {author} {\bibfnamefont {G.}~\bibnamefont {Bertone}},\
  }\href@noop {} {\bibfield  {journal} {\bibinfo  {journal} {Journal of
  Cosmology and Astroparticle Physics}\ }\textbf {\bibinfo {volume} {2013}},\
  \bibinfo {pages} {014} (\bibinfo {year} {2013})}\BibitemShut {NoStop}%
\bibitem [{\citenamefont {Polisensky}\ and\ \citenamefont
  {Ricotti}(2011)}]{Polisensky:2010rw}%
  \BibitemOpen
  \bibfield  {author} {\bibinfo {author} {\bibfnamefont {E.}~\bibnamefont
  {Polisensky}}\ and\ \bibinfo {author} {\bibfnamefont {M.}~\bibnamefont
  {Ricotti}},\ }\href {\doibase 10.1103/PhysRevD.83.043506} {\bibfield
  {journal} {\bibinfo  {journal} {Phys. Rev. D}\ }\textbf {\bibinfo {volume}
  {83}},\ \bibinfo {pages} {043506} (\bibinfo {year} {2011})},\ \Eprint
  {http://arxiv.org/abs/1004.1459} {arXiv:1004.1459 [astro-ph.CO]} \BibitemShut
  {NoStop}%
\bibitem [{\citenamefont {Cherry}\ and\ \citenamefont
  {Horiuchi}(2017)}]{Cherry:2017dwu}%
  \BibitemOpen
  \bibfield  {author} {\bibinfo {author} {\bibfnamefont {J.~F.}\ \bibnamefont
  {Cherry}}\ and\ \bibinfo {author} {\bibfnamefont {S.}~\bibnamefont
  {Horiuchi}},\ }\href {\doibase 10.1103/PhysRevD.95.083015} {\bibfield
  {journal} {\bibinfo  {journal} {Phys. Rev. D}\ }\textbf {\bibinfo {volume}
  {95}},\ \bibinfo {pages} {083015} (\bibinfo {year} {2017})},\ \Eprint
  {http://arxiv.org/abs/1701.07874} {arXiv:1701.07874 [hep-ph]} \BibitemShut
  {NoStop}%
\bibitem [{\citenamefont {Nadler}\ \emph {et~al.}(2019)\citenamefont {Nadler},
  \citenamefont {Gluscevic}, \citenamefont {Boddy},\ and\ \citenamefont
  {Wechsler}}]{Nadler:2019zrb}%
  \BibitemOpen
  \bibfield  {author} {\bibinfo {author} {\bibfnamefont {E.~O.}\ \bibnamefont
  {Nadler}}, \bibinfo {author} {\bibfnamefont {V.}~\bibnamefont {Gluscevic}},
  \bibinfo {author} {\bibfnamefont {K.~K.}\ \bibnamefont {Boddy}}, \ and\
  \bibinfo {author} {\bibfnamefont {R.~H.}\ \bibnamefont {Wechsler}},\ }\href
  {\doibase 10.3847/2041-8213/ab1eb2} {\bibfield  {journal} {\bibinfo
  {journal} {Astrophys. J. Lett.}\ }\textbf {\bibinfo {volume} {878}},\
  \bibinfo {pages} {32} (\bibinfo {year} {2019})},\ \bibinfo {note} {[Erratum:
  Astrophys.J.Lett. 897, L46 (2020), Erratum: Astrophys.J. 897, L46 (2020)]},\
  \Eprint {http://arxiv.org/abs/1904.10000} {arXiv:1904.10000 [astro-ph.CO]}
  \BibitemShut {NoStop}%
\bibitem [{\citenamefont {Gilman}\ \emph {et~al.}(2020)\citenamefont {Gilman},
  \citenamefont {Birrer}, \citenamefont {Nierenberg}, \citenamefont {Treu},
  \citenamefont {Du},\ and\ \citenamefont {Benson}}]{Gilman:2019nap}%
  \BibitemOpen
  \bibfield  {author} {\bibinfo {author} {\bibfnamefont {D.}~\bibnamefont
  {Gilman}}, \bibinfo {author} {\bibfnamefont {S.}~\bibnamefont {Birrer}},
  \bibinfo {author} {\bibfnamefont {A.}~\bibnamefont {Nierenberg}}, \bibinfo
  {author} {\bibfnamefont {T.}~\bibnamefont {Treu}}, \bibinfo {author}
  {\bibfnamefont {X.}~\bibnamefont {Du}}, \ and\ \bibinfo {author}
  {\bibfnamefont {A.}~\bibnamefont {Benson}},\ }\href {\doibase
  10.1093/mnras/stz3480} {\bibfield  {journal} {\bibinfo  {journal} {Mon. Not.
  Roy. Astron. Soc.}\ }\textbf {\bibinfo {volume} {491}},\ \bibinfo {pages}
  {6077} (\bibinfo {year} {2020})},\ \Eprint {http://arxiv.org/abs/1908.06983}
  {arXiv:1908.06983 [astro-ph.CO]} \BibitemShut {NoStop}%
\bibitem [{\citenamefont {Banik}\ \emph {et~al.}(2021)\citenamefont {Banik},
  \citenamefont {Bovy}, \citenamefont {Bertone}, \citenamefont {Erkal},\ and\
  \citenamefont {de~Boer}}]{Banik:2019smi}%
  \BibitemOpen
  \bibfield  {author} {\bibinfo {author} {\bibfnamefont {N.}~\bibnamefont
  {Banik}}, \bibinfo {author} {\bibfnamefont {J.}~\bibnamefont {Bovy}},
  \bibinfo {author} {\bibfnamefont {G.}~\bibnamefont {Bertone}}, \bibinfo
  {author} {\bibfnamefont {D.}~\bibnamefont {Erkal}}, \ and\ \bibinfo {author}
  {\bibfnamefont {T.~J.~L.}\ \bibnamefont {de~Boer}},\ }\href {\doibase
  10.1088/1475-7516/2021/10/043} {\bibfield  {journal} {\bibinfo  {journal}
  {JCAP}\ }\textbf {\bibinfo {volume} {10}},\ \bibinfo {pages} {043} (\bibinfo
  {year} {2021})},\ \Eprint {http://arxiv.org/abs/1911.02663} {arXiv:1911.02663
  [astro-ph.GA]} \BibitemShut {NoStop}%
\bibitem [{\citenamefont {Ir\v{s}i\v{c}}\ \emph {et~al.}(2017)\citenamefont
  {Ir\v{s}i\v{c}} \emph {et~al.}}]{Irsic:2017ixq}%
  \BibitemOpen
  \bibfield  {author} {\bibinfo {author} {\bibfnamefont {V.}~\bibnamefont
  {Ir\v{s}i\v{c}}} \emph {et~al.},\ }\href {\doibase
  10.1103/PhysRevD.96.023522} {\bibfield  {journal} {\bibinfo  {journal} {Phys.
  Rev. D}\ }\textbf {\bibinfo {volume} {96}},\ \bibinfo {pages} {023522}
  (\bibinfo {year} {2017})},\ \Eprint {http://arxiv.org/abs/1702.01764}
  {arXiv:1702.01764 [astro-ph.CO]} \BibitemShut {NoStop}%
\bibitem [{\citenamefont {Nadler}\ \emph {et~al.}(2021)\citenamefont {Nadler},
  \citenamefont {Birrer}, \citenamefont {Gilman}, \citenamefont {Wechsler},
  \citenamefont {Du}, \citenamefont {Benson}, \citenamefont {Nierenberg},\ and\
  \citenamefont {Treu}}]{Nadler:2021dft}%
  \BibitemOpen
  \bibfield  {author} {\bibinfo {author} {\bibfnamefont {E.~O.}\ \bibnamefont
  {Nadler}}, \bibinfo {author} {\bibfnamefont {S.}~\bibnamefont {Birrer}},
  \bibinfo {author} {\bibfnamefont {D.}~\bibnamefont {Gilman}}, \bibinfo
  {author} {\bibfnamefont {R.~H.}\ \bibnamefont {Wechsler}}, \bibinfo {author}
  {\bibfnamefont {X.}~\bibnamefont {Du}}, \bibinfo {author} {\bibfnamefont
  {A.}~\bibnamefont {Benson}}, \bibinfo {author} {\bibfnamefont {A.~M.}\
  \bibnamefont {Nierenberg}}, \ and\ \bibinfo {author} {\bibfnamefont
  {T.}~\bibnamefont {Treu}},\ }\href {\doibase 10.3847/1538-4357/abf9a3}
  {\bibfield  {journal} {\bibinfo  {journal} {Astrophys. J.}\ }\textbf
  {\bibinfo {volume} {917}},\ \bibinfo {pages} {7} (\bibinfo {year} {2021})},\
  \Eprint {http://arxiv.org/abs/2101.07810} {arXiv:2101.07810 [astro-ph.CO]}
  \BibitemShut {NoStop}%
\bibitem [{\citenamefont {Gilman}\ \emph {et~al.}(2021)\citenamefont {Gilman},
  \citenamefont {Bovy}, \citenamefont {Treu}, \citenamefont {Nierenberg},
  \citenamefont {Birrer}, \citenamefont {Benson},\ and\ \citenamefont
  {Sameie}}]{Gilman:2021sdr}%
  \BibitemOpen
  \bibfield  {author} {\bibinfo {author} {\bibfnamefont {D.}~\bibnamefont
  {Gilman}}, \bibinfo {author} {\bibfnamefont {J.}~\bibnamefont {Bovy}},
  \bibinfo {author} {\bibfnamefont {T.}~\bibnamefont {Treu}}, \bibinfo {author}
  {\bibfnamefont {A.}~\bibnamefont {Nierenberg}}, \bibinfo {author}
  {\bibfnamefont {S.}~\bibnamefont {Birrer}}, \bibinfo {author} {\bibfnamefont
  {A.}~\bibnamefont {Benson}}, \ and\ \bibinfo {author} {\bibfnamefont
  {O.}~\bibnamefont {Sameie}},\ }\href {\doibase 10.1093/mnras/stab2335}
  {\bibfield  {journal} {\bibinfo  {journal} {Mon. Not. Roy. Astron. Soc.}\
  }\textbf {\bibinfo {volume} {507}},\ \bibinfo {pages} {2432} (\bibinfo {year}
  {2021})},\ \Eprint {http://arxiv.org/abs/2105.05259} {arXiv:2105.05259
  [astro-ph.CO]} \BibitemShut {NoStop}%
\bibitem [{\citenamefont {Zelko}\ \emph {et~al.}(2022)\citenamefont {Zelko},
  \citenamefont {Treu}, \citenamefont {Abazajian}, \citenamefont {Gilman},
  \citenamefont {Benson}, \citenamefont {Birrer}, \citenamefont {Nierenberg},\
  and\ \citenamefont {Kusenko}}]{Zelko:2022tgf}%
  \BibitemOpen
  \bibfield  {author} {\bibinfo {author} {\bibfnamefont {I.~A.}\ \bibnamefont
  {Zelko}}, \bibinfo {author} {\bibfnamefont {T.}~\bibnamefont {Treu}},
  \bibinfo {author} {\bibfnamefont {K.~N.}\ \bibnamefont {Abazajian}}, \bibinfo
  {author} {\bibfnamefont {D.}~\bibnamefont {Gilman}}, \bibinfo {author}
  {\bibfnamefont {A.~J.}\ \bibnamefont {Benson}}, \bibinfo {author}
  {\bibfnamefont {S.}~\bibnamefont {Birrer}}, \bibinfo {author} {\bibfnamefont
  {A.~M.}\ \bibnamefont {Nierenberg}}, \ and\ \bibinfo {author} {\bibfnamefont
  {A.}~\bibnamefont {Kusenko}},\ }\href@noop {} {\  (\bibinfo {year} {2022})},\
  \Eprint {http://arxiv.org/abs/2205.09777} {arXiv:2205.09777 [hep-ph]}
  \BibitemShut {NoStop}%
\bibitem [{\citenamefont {Abazajian}\ and\ \citenamefont
  {Koushiappas}(2006)}]{Abazajian:2006yn}%
  \BibitemOpen
  \bibfield  {author} {\bibinfo {author} {\bibfnamefont {K.}~\bibnamefont
  {Abazajian}}\ and\ \bibinfo {author} {\bibfnamefont {S.~M.}\ \bibnamefont
  {Koushiappas}},\ }\href {\doibase 10.1103/PhysRevD.74.023527} {\bibfield
  {journal} {\bibinfo  {journal} {Phys. Rev. D}\ }\textbf {\bibinfo {volume}
  {74}},\ \bibinfo {pages} {023527} (\bibinfo {year} {2006})},\ \Eprint
  {http://arxiv.org/abs/astro-ph/0605271} {arXiv:astro-ph/0605271} \BibitemShut
  {NoStop}%
\bibitem [{\citenamefont {Nierenberg}\ \emph {et~al.}(2021)\citenamefont
  {Nierenberg}, \citenamefont {Bennert}, \citenamefont {Benson}, \citenamefont
  {Birrer}, \citenamefont {Djorgovski}, \citenamefont {Du}, \citenamefont
  {Fassnacht}, \citenamefont {Gilman}, \citenamefont {Hoenig}, \citenamefont
  {Kusenko} \emph {et~al.}}]{nierenberg2021definitive}%
  \BibitemOpen
  \bibfield  {author} {\bibinfo {author} {\bibfnamefont {A.}~\bibnamefont
  {Nierenberg}}, \bibinfo {author} {\bibfnamefont {V.~N.}\ \bibnamefont
  {Bennert}}, \bibinfo {author} {\bibfnamefont {A.}~\bibnamefont {Benson}},
  \bibinfo {author} {\bibfnamefont {S.}~\bibnamefont {Birrer}}, \bibinfo
  {author} {\bibfnamefont {S.~G.}\ \bibnamefont {Djorgovski}}, \bibinfo
  {author} {\bibfnamefont {X.}~\bibnamefont {Du}}, \bibinfo {author}
  {\bibfnamefont {C.}~\bibnamefont {Fassnacht}}, \bibinfo {author}
  {\bibfnamefont {D.}~\bibnamefont {Gilman}}, \bibinfo {author} {\bibfnamefont
  {S.~F.}\ \bibnamefont {Hoenig}}, \bibinfo {author} {\bibfnamefont
  {A.}~\bibnamefont {Kusenko}},  \emph {et~al.},\ }\href@noop {} {\bibfield
  {journal} {\bibinfo  {journal} {JWST Proposal. Cycle 1}\ ,\ \bibinfo {pages}
  {2046}} (\bibinfo {year} {2021})}\BibitemShut {NoStop}%
\bibitem [{\citenamefont {Moustakas}\ \emph {et~al.}(2009)\citenamefont
  {Moustakas} \emph {et~al.}}]{Moustakas:2009na}%
  \BibitemOpen
  \bibfield  {author} {\bibinfo {author} {\bibfnamefont {L.~A.}\ \bibnamefont
  {Moustakas}} \emph {et~al.},\ }\href@noop {} {\  (\bibinfo {year} {2009})},\
  \Eprint {http://arxiv.org/abs/0902.3219} {arXiv:0902.3219 [astro-ph.CO]}
  \BibitemShut {NoStop}%
\bibitem [{\citenamefont {Pagels}\ and\ \citenamefont
  {Primack}(1982)}]{pagels1982grav}%
  \BibitemOpen
  \bibfield  {author} {\bibinfo {author} {\bibfnamefont {H.}~\bibnamefont
  {Pagels}}\ and\ \bibinfo {author} {\bibfnamefont {J.~R.}\ \bibnamefont
  {Primack}},\ }\href@noop {} {\bibfield  {journal} {\bibinfo  {journal}
  {Physical Review Letters}\ }\textbf {\bibinfo {volume} {48}},\ \bibinfo
  {pages} {223} (\bibinfo {year} {1982})}\BibitemShut {NoStop}%
\bibitem [{\citenamefont {Tseliakhovich}\ and\ \citenamefont
  {Hirata}(2010)}]{Tseliakhovich:2010bj}%
  \BibitemOpen
  \bibfield  {author} {\bibinfo {author} {\bibfnamefont {D.}~\bibnamefont
  {Tseliakhovich}}\ and\ \bibinfo {author} {\bibfnamefont {C.}~\bibnamefont
  {Hirata}},\ }\href {\doibase 10.1103/PhysRevD.82.083520} {\bibfield
  {journal} {\bibinfo  {journal} {Phys. Rev. D}\ }\textbf {\bibinfo {volume}
  {82}},\ \bibinfo {pages} {083520} (\bibinfo {year} {2010})},\ \Eprint
  {http://arxiv.org/abs/1005.2416} {arXiv:1005.2416 [astro-ph.CO]} \BibitemShut
  {NoStop}%
\bibitem [{\citenamefont {Naoz}\ \emph {et~al.}(2006)\citenamefont {Naoz},
  \citenamefont {Noter},\ and\ \citenamefont {Barkana}}]{Naoz:2006tr}%
  \BibitemOpen
  \bibfield  {author} {\bibinfo {author} {\bibfnamefont {S.}~\bibnamefont
  {Naoz}}, \bibinfo {author} {\bibfnamefont {S.}~\bibnamefont {Noter}}, \ and\
  \bibinfo {author} {\bibfnamefont {R.}~\bibnamefont {Barkana}},\ }\href
  {\doibase 10.1111/j.1745-3933.2006.00251.x} {\bibfield  {journal} {\bibinfo
  {journal} {Mon. Not. Roy. Astron. Soc.}\ }\textbf {\bibinfo {volume} {373}},\
  \bibinfo {pages} {L98} (\bibinfo {year} {2006})},\ \Eprint
  {http://arxiv.org/abs/astro-ph/0604050} {arXiv:astro-ph/0604050} \BibitemShut
  {NoStop}%
\bibitem [{\citenamefont {Naoz}\ and\ \citenamefont
  {Barkana}(2007)}]{Naoz:2007fo}%
  \BibitemOpen
  \bibfield  {author} {\bibinfo {author} {\bibfnamefont {S.}~\bibnamefont
  {Naoz}}\ and\ \bibinfo {author} {\bibfnamefont {R.}~\bibnamefont {Barkana}},\
  }\href@noop {} {\bibfield  {journal} {\bibinfo  {journal} {Monthly Notices of
  the Royal Astronomical Society}\ }\textbf {\bibinfo {volume} {377}},\
  \bibinfo {pages} {667} (\bibinfo {year} {2007})}\BibitemShut {NoStop}%
\bibitem [{\citenamefont {Oh}\ and\ \citenamefont {Haiman}(2002)}]{Oh:2002se}%
  \BibitemOpen
  \bibfield  {author} {\bibinfo {author} {\bibfnamefont {S.~P.}\ \bibnamefont
  {Oh}}\ and\ \bibinfo {author} {\bibfnamefont {Z.}~\bibnamefont {Haiman}},\
  }\href@noop {} {\bibfield  {journal} {\bibinfo  {journal} {The Astrophysical
  Journal}\ }\textbf {\bibinfo {volume} {569}},\ \bibinfo {pages} {558}
  (\bibinfo {year} {2002})}\BibitemShut {NoStop}%
\bibitem [{\citenamefont {Hirata}(2018)}]{Hirata:2017ivs}%
  \BibitemOpen
  \bibfield  {author} {\bibinfo {author} {\bibfnamefont {C.~M.}\ \bibnamefont
  {Hirata}},\ }\href {\doibase 10.1093/mnras/stx2854} {\bibfield  {journal}
  {\bibinfo  {journal} {Mon. Not. Roy. Astron. Soc.}\ }\textbf {\bibinfo
  {volume} {474}},\ \bibinfo {pages} {2173} (\bibinfo {year} {2018})},\ \Eprint
  {http://arxiv.org/abs/1707.03358} {arXiv:1707.03358 [astro-ph.CO]}
  \BibitemShut {NoStop}%
\bibitem [{\citenamefont {Dodelson}\ and\ \citenamefont
  {Widrow}(1994)}]{dodelson1994sterile}%
  \BibitemOpen
  \bibfield  {author} {\bibinfo {author} {\bibfnamefont {S.}~\bibnamefont
  {Dodelson}}\ and\ \bibinfo {author} {\bibfnamefont {L.~M.}\ \bibnamefont
  {Widrow}},\ }\href@noop {} {\bibfield  {journal} {\bibinfo  {journal}
  {Physical Review Letters}\ }\textbf {\bibinfo {volume} {72}},\ \bibinfo
  {pages} {17} (\bibinfo {year} {1994})}\BibitemShut {NoStop}%
\bibitem [{\citenamefont {Shi}\ and\ \citenamefont
  {Fuller}(1999)}]{shi1999new}%
  \BibitemOpen
  \bibfield  {author} {\bibinfo {author} {\bibfnamefont {X.}~\bibnamefont
  {Shi}}\ and\ \bibinfo {author} {\bibfnamefont {G.~M.}\ \bibnamefont
  {Fuller}},\ }\href@noop {} {\bibfield  {journal} {\bibinfo  {journal}
  {Physical Review Letters}\ }\textbf {\bibinfo {volume} {82}},\ \bibinfo
  {pages} {2832} (\bibinfo {year} {1999})}\BibitemShut {NoStop}%
\bibitem [{\citenamefont {Rajagopal}\ \emph {et~al.}(1991)\citenamefont
  {Rajagopal}, \citenamefont {Turner},\ and\ \citenamefont
  {Wilczek}}]{wilczek1991cosmological}%
  \BibitemOpen
  \bibfield  {author} {\bibinfo {author} {\bibfnamefont {K.}~\bibnamefont
  {Rajagopal}}, \bibinfo {author} {\bibfnamefont {M.~S.}\ \bibnamefont
  {Turner}}, \ and\ \bibinfo {author} {\bibfnamefont {F.}~\bibnamefont
  {Wilczek}},\ }\href@noop {} {\bibfield  {journal} {\bibinfo  {journal}
  {Nuclear Physics B}\ }\textbf {\bibinfo {volume} {358}},\ \bibinfo {pages}
  {447} (\bibinfo {year} {1991})}\BibitemShut {NoStop}%
\bibitem [{\citenamefont {Chun}\ \emph {et~al.}(1992)\citenamefont {Chun},
  \citenamefont {Kim},\ and\ \citenamefont {Nilles}}]{chun1992axino}%
  \BibitemOpen
  \bibfield  {author} {\bibinfo {author} {\bibfnamefont {E.}~\bibnamefont
  {Chun}}, \bibinfo {author} {\bibfnamefont {J.~E.}\ \bibnamefont {Kim}}, \
  and\ \bibinfo {author} {\bibfnamefont {H.~P.}\ \bibnamefont {Nilles}},\
  }\href@noop {} {\bibfield  {journal} {\bibinfo  {journal} {Physics Letters
  B}\ }\textbf {\bibinfo {volume} {287}},\ \bibinfo {pages} {123} (\bibinfo
  {year} {1992})}\BibitemShut {NoStop}%
\bibitem [{\citenamefont {Kim}\ and\ \citenamefont
  {Seo}(2012)}]{kim2012mixing}%
  \BibitemOpen
  \bibfield  {author} {\bibinfo {author} {\bibfnamefont {J.~E.}\ \bibnamefont
  {Kim}}\ and\ \bibinfo {author} {\bibfnamefont {M.-S.}\ \bibnamefont {Seo}},\
  }\href@noop {} {\bibfield  {journal} {\bibinfo  {journal} {Nuclear Physics
  B}\ }\textbf {\bibinfo {volume} {864}},\ \bibinfo {pages} {296} (\bibinfo
  {year} {2012})}\BibitemShut {NoStop}%
\bibitem [{\citenamefont {Jaramillo}(2022)}]{Jaramillo:2022mos}%
  \BibitemOpen
  \bibfield  {author} {\bibinfo {author} {\bibfnamefont {C.}~\bibnamefont
  {Jaramillo}},\ }\href {\doibase 10.1088/1475-7516/2022/10/093} {\bibfield
  {journal} {\bibinfo  {journal} {JCAP}\ }\textbf {\bibinfo {volume} {10}},\
  \bibinfo {pages} {093} (\bibinfo {year} {2022})},\ \Eprint
  {http://arxiv.org/abs/2207.11269} {arXiv:2207.11269 [hep-ph]} \BibitemShut
  {NoStop}%
\bibitem [{\citenamefont {Fayet}(1977)}]{fayet1977mixing}%
  \BibitemOpen
  \bibfield  {author} {\bibinfo {author} {\bibfnamefont {P.}~\bibnamefont
  {Fayet}},\ }\href@noop {} {\bibfield  {journal} {\bibinfo  {journal} {Physics
  Letters B}\ }\textbf {\bibinfo {volume} {70}},\ \bibinfo {pages} {461}
  (\bibinfo {year} {1977})}\BibitemShut {NoStop}%
\bibitem [{\citenamefont {Choi}\ \emph {et~al.}(1999)\citenamefont {Choi},
  \citenamefont {Hwang}, \citenamefont {Kim},\ and\ \citenamefont
  {Lee}}]{choi1999cosmological}%
  \BibitemOpen
  \bibfield  {author} {\bibinfo {author} {\bibfnamefont {K.}~\bibnamefont
  {Choi}}, \bibinfo {author} {\bibfnamefont {K.}~\bibnamefont {Hwang}},
  \bibinfo {author} {\bibfnamefont {H.~B.}\ \bibnamefont {Kim}}, \ and\
  \bibinfo {author} {\bibfnamefont {T.}~\bibnamefont {Lee}},\ }\href@noop {}
  {\bibfield  {journal} {\bibinfo  {journal} {Physics Letters B}\ }\textbf
  {\bibinfo {volume} {467}},\ \bibinfo {pages} {211} (\bibinfo {year}
  {1999})}\BibitemShut {NoStop}%
\bibitem [{\citenamefont {Bolz}\ \emph {et~al.}(2001)\citenamefont {Bolz},
  \citenamefont {Brandenburg},\ and\ \citenamefont
  {Buchm{\"u}ller}}]{bolz2001thermal}%
  \BibitemOpen
  \bibfield  {author} {\bibinfo {author} {\bibfnamefont {M.}~\bibnamefont
  {Bolz}}, \bibinfo {author} {\bibfnamefont {A.}~\bibnamefont {Brandenburg}}, \
  and\ \bibinfo {author} {\bibfnamefont {W.}~\bibnamefont {Buchm{\"u}ller}},\
  }\href@noop {} {\bibfield  {journal} {\bibinfo  {journal} {Nuclear Physics
  B}\ }\textbf {\bibinfo {volume} {606}},\ \bibinfo {pages} {518} (\bibinfo
  {year} {2001})}\BibitemShut {NoStop}%
\bibitem [{\citenamefont {Casalbuoni}\ \emph {et~al.}(1988)\citenamefont
  {Casalbuoni}, \citenamefont {De~Curtis}, \citenamefont {Dominici},
  \citenamefont {Feruglio},\ and\ \citenamefont
  {Gatto}}]{casalbuoni1988gravitino}%
  \BibitemOpen
  \bibfield  {author} {\bibinfo {author} {\bibfnamefont {R.}~\bibnamefont
  {Casalbuoni}}, \bibinfo {author} {\bibfnamefont {S.}~\bibnamefont
  {De~Curtis}}, \bibinfo {author} {\bibfnamefont {D.}~\bibnamefont {Dominici}},
  \bibinfo {author} {\bibfnamefont {F.}~\bibnamefont {Feruglio}}, \ and\
  \bibinfo {author} {\bibfnamefont {R.}~\bibnamefont {Gatto}},\ }\href@noop {}
  {\bibfield  {journal} {\bibinfo  {journal} {Physics Letters B}\ }\textbf
  {\bibinfo {volume} {215}},\ \bibinfo {pages} {313} (\bibinfo {year}
  {1988})}\BibitemShut {NoStop}%
\bibitem [{\citenamefont {Garcia}\ \emph {et~al.}(2020)\citenamefont {Garcia},
  \citenamefont {Mambrini}, \citenamefont {Olive},\ and\ \citenamefont
  {Verner}}]{Garcia:2020hyo}%
  \BibitemOpen
  \bibfield  {author} {\bibinfo {author} {\bibfnamefont {M.~A.~G.}\
  \bibnamefont {Garcia}}, \bibinfo {author} {\bibfnamefont {Y.}~\bibnamefont
  {Mambrini}}, \bibinfo {author} {\bibfnamefont {K.~A.}\ \bibnamefont {Olive}},
  \ and\ \bibinfo {author} {\bibfnamefont {S.}~\bibnamefont {Verner}},\ }\href
  {\doibase 10.1103/PhysRevD.102.083533} {\bibfield  {journal} {\bibinfo
  {journal} {Phys. Rev. D}\ }\textbf {\bibinfo {volume} {102}},\ \bibinfo
  {pages} {083533} (\bibinfo {year} {2020})},\ \Eprint
  {http://arxiv.org/abs/2006.03325} {arXiv:2006.03325 [hep-ph]} \BibitemShut
  {NoStop}%
\bibitem [{\citenamefont {Ballesteros}\ \emph {et~al.}(2021)\citenamefont
  {Ballesteros}, \citenamefont {Garcia},\ and\ \citenamefont
  {Pierre}}]{Ballesteros:2020adh}%
  \BibitemOpen
  \bibfield  {author} {\bibinfo {author} {\bibfnamefont {G.}~\bibnamefont
  {Ballesteros}}, \bibinfo {author} {\bibfnamefont {M.~A.~G.}\ \bibnamefont
  {Garcia}}, \ and\ \bibinfo {author} {\bibfnamefont {M.}~\bibnamefont
  {Pierre}},\ }\href {\doibase 10.1088/1475-7516/2021/03/101} {\bibfield
  {journal} {\bibinfo  {journal} {JCAP}\ }\textbf {\bibinfo {volume} {03}},\
  \bibinfo {pages} {101} (\bibinfo {year} {2021})},\ \Eprint
  {http://arxiv.org/abs/2011.13458} {arXiv:2011.13458 [hep-ph]} \BibitemShut
  {NoStop}%
\bibitem [{\citenamefont {Viel}\ \emph {et~al.}(2005)\citenamefont {Viel},
  \citenamefont {Lesgourgues}, \citenamefont {Haehnelt}, \citenamefont
  {Matarrese},\ and\ \citenamefont {Riotto}}]{viel2005}%
  \BibitemOpen
  \bibfield  {author} {\bibinfo {author} {\bibfnamefont {M.}~\bibnamefont
  {Viel}}, \bibinfo {author} {\bibfnamefont {J.}~\bibnamefont {Lesgourgues}},
  \bibinfo {author} {\bibfnamefont {M.~G.}\ \bibnamefont {Haehnelt}}, \bibinfo
  {author} {\bibfnamefont {S.}~\bibnamefont {Matarrese}}, \ and\ \bibinfo
  {author} {\bibfnamefont {A.}~\bibnamefont {Riotto}},\ }\href@noop {}
  {\bibfield  {journal} {\bibinfo  {journal} {Physical Review D}\ }\textbf
  {\bibinfo {volume} {71}},\ \bibinfo {pages} {063534} (\bibinfo {year}
  {2005})}\BibitemShut {NoStop}%
\bibitem [{\citenamefont {Baltz}\ and\ \citenamefont
  {Murayama}(2003)}]{baltz2003gravitino}%
  \BibitemOpen
  \bibfield  {author} {\bibinfo {author} {\bibfnamefont {E.~A.}\ \bibnamefont
  {Baltz}}\ and\ \bibinfo {author} {\bibfnamefont {H.}~\bibnamefont
  {Murayama}},\ }\href@noop {} {\bibfield  {journal} {\bibinfo  {journal}
  {Journal of High Energy Physics}\ }\textbf {\bibinfo {volume} {2003}},\
  \bibinfo {pages} {067} (\bibinfo {year} {2003})}\BibitemShut {NoStop}%
\bibitem [{\citenamefont {Viel}\ \emph {et~al.}(2012)\citenamefont {Viel},
  \citenamefont {Markovi{\v{c}}}, \citenamefont {Baldi},\ and\ \citenamefont
  {Weller}}]{viel2012non}%
  \BibitemOpen
  \bibfield  {author} {\bibinfo {author} {\bibfnamefont {M.}~\bibnamefont
  {Viel}}, \bibinfo {author} {\bibfnamefont {K.}~\bibnamefont
  {Markovi{\v{c}}}}, \bibinfo {author} {\bibfnamefont {M.}~\bibnamefont
  {Baldi}}, \ and\ \bibinfo {author} {\bibfnamefont {J.}~\bibnamefont
  {Weller}},\ }\href@noop {} {\bibfield  {journal} {\bibinfo  {journal}
  {Monthly Notices of the Royal Astronomical Society}\ }\textbf {\bibinfo
  {volume} {421}},\ \bibinfo {pages} {50} (\bibinfo {year} {2012})}\BibitemShut
  {NoStop}%
\bibitem [{\citenamefont {Hansen}\ \emph {et~al.}(2002)\citenamefont {Hansen},
  \citenamefont {Lesgourgues}, \citenamefont {Pastor},\ and\ \citenamefont
  {Silk}}]{hansen2002constraining}%
  \BibitemOpen
  \bibfield  {author} {\bibinfo {author} {\bibfnamefont {S.~H.}\ \bibnamefont
  {Hansen}}, \bibinfo {author} {\bibfnamefont {J.}~\bibnamefont {Lesgourgues}},
  \bibinfo {author} {\bibfnamefont {S.}~\bibnamefont {Pastor}}, \ and\ \bibinfo
  {author} {\bibfnamefont {J.}~\bibnamefont {Silk}},\ }\href@noop {} {\bibfield
   {journal} {\bibinfo  {journal} {Monthly Notices of the Royal Astronomical
  Society}\ }\textbf {\bibinfo {volume} {333}},\ \bibinfo {pages} {544}
  (\bibinfo {year} {2002})}\BibitemShut {NoStop}%
\bibitem [{\citenamefont
  {Abazajian}(2006{\natexlab{b}})}]{abazajian2006production}%
  \BibitemOpen
  \bibfield  {author} {\bibinfo {author} {\bibfnamefont {K.}~\bibnamefont
  {Abazajian}},\ }\href@noop {} {\bibfield  {journal} {\bibinfo  {journal}
  {Physical Review D}\ }\textbf {\bibinfo {volume} {73}},\ \bibinfo {pages}
  {063506} (\bibinfo {year} {2006}{\natexlab{b}})}\BibitemShut {NoStop}%
\bibitem [{\citenamefont {Blas}\ \emph {et~al.}(2011)\citenamefont {Blas},
  \citenamefont {Lesgourgues},\ and\ \citenamefont {Tram}}]{blas2011cosmic}%
  \BibitemOpen
  \bibfield  {author} {\bibinfo {author} {\bibfnamefont {D.}~\bibnamefont
  {Blas}}, \bibinfo {author} {\bibfnamefont {J.}~\bibnamefont {Lesgourgues}}, \
  and\ \bibinfo {author} {\bibfnamefont {T.}~\bibnamefont {Tram}},\ }\href@noop
  {} {\bibfield  {journal} {\bibinfo  {journal} {Journal of Cosmology and
  Astroparticle Physics}\ }\textbf {\bibinfo {volume} {2011}},\ \bibinfo
  {pages} {034} (\bibinfo {year} {2011})}\BibitemShut {NoStop}%
\bibitem [{\citenamefont {Aghanim}\ \emph {et~al.}(2020)\citenamefont
  {Aghanim}, \citenamefont {Akrami}, \citenamefont {Ashdown}, \citenamefont
  {Aumont}, \citenamefont {Baccigalupi}, \citenamefont {Ballardini},
  \citenamefont {Banday}, \citenamefont {Barreiro}, \citenamefont {Bartolo},
  \citenamefont {Basak} \emph {et~al.}}]{aghanim2020planck}%
  \BibitemOpen
  \bibfield  {author} {\bibinfo {author} {\bibfnamefont {N.}~\bibnamefont
  {Aghanim}}, \bibinfo {author} {\bibfnamefont {Y.}~\bibnamefont {Akrami}},
  \bibinfo {author} {\bibfnamefont {M.}~\bibnamefont {Ashdown}}, \bibinfo
  {author} {\bibfnamefont {J.}~\bibnamefont {Aumont}}, \bibinfo {author}
  {\bibfnamefont {C.}~\bibnamefont {Baccigalupi}}, \bibinfo {author}
  {\bibfnamefont {M.}~\bibnamefont {Ballardini}}, \bibinfo {author}
  {\bibfnamefont {A.}~\bibnamefont {Banday}}, \bibinfo {author} {\bibfnamefont
  {R.}~\bibnamefont {Barreiro}}, \bibinfo {author} {\bibfnamefont
  {N.}~\bibnamefont {Bartolo}}, \bibinfo {author} {\bibfnamefont
  {S.}~\bibnamefont {Basak}},  \emph {et~al.},\ }\href@noop {} {\bibfield
  {journal} {\bibinfo  {journal} {Astronomy \& Astrophysics}\ }\textbf
  {\bibinfo {volume} {641}},\ \bibinfo {pages} {A6} (\bibinfo {year}
  {2020})}\BibitemShut {NoStop}%
\bibitem [{\citenamefont {Villasenor}\ \emph {et~al.}(2022)\citenamefont
  {Villasenor}, \citenamefont {Robertson}, \citenamefont {Madau},\ and\
  \citenamefont {Schneider}}]{Villasenor:2022aiy}%
  \BibitemOpen
  \bibfield  {author} {\bibinfo {author} {\bibfnamefont {B.}~\bibnamefont
  {Villasenor}}, \bibinfo {author} {\bibfnamefont {B.}~\bibnamefont
  {Robertson}}, \bibinfo {author} {\bibfnamefont {P.}~\bibnamefont {Madau}}, \
  and\ \bibinfo {author} {\bibfnamefont {E.}~\bibnamefont {Schneider}},\
  }\href@noop {} {\  (\bibinfo {year} {2022})},\ \Eprint
  {http://arxiv.org/abs/2209.14220} {arXiv:2209.14220 [astro-ph.CO]}
  \BibitemShut {NoStop}%
\end{thebibliography}%
\bibliographystyle{apsrev4-1}
\end{document}